# Multivariable theory of droplet nucleation in a single-component vapor

Nikolay V. Alekseechkin

Akhiezer Institute for Theoretical Physics, National Science Centre "Kharkov Institute of Physics and Technology", Akademicheskaya Street 1, Kharkov 61108, Ukraine
Email: n.alex@kipt.kharkov.ua

**Abstract**. The multivariable theory of nucleation [N. V. Alekseechkin, J. Chem. Phys. **124**, 124512 (2006)] is applied to the droplet nucleation in a supersaturated single-component vapor; the droplet volume $V$, temperature $T$, and volume change rate $U = \dot{V}$ are the variables of the theory. A new approach based on macroscopic kinetics is developed for the droplet evolution and results in the derived equations for $\dot{U}$, $\dot{V}$, and $\dot{T}$. It is shown that there is no the viscosity effect in the employed ideal gas approximation, so the variable $U$ can be omitted. The nonisothermal effect (the discrepancy between the actual and isothermal nucleation rates) earlier studied numerically is analytically examined here. The calculated steady state distribution function of droplets shows their average overheating relatively the vapor temperature. An inert background gas is shown to diminish the nonisothermal effect in comparison with a pure vapor case.

## 1. Introduction

Calculation of the nucleation rate of liquid droplets in the process of condensation of a supersaturated vapor [1-14] is the classical problem of the nucleation theory which starts from the thirties of last century [1-4] and still topical. In the case of a single-component vapor, the employed theory, as a rule, is (i) one-dimensional and (ii) based on the standard (*microscopic*) pattern of evolution of the new phase embryo: the change of its size occurs due to the attachment (with the probability $w_+$ per unit time per unit area) and detachment (with the probability $w_-$) of monomers. The probabilities $w_+$ and $w_-$ are related by the detailed balance

$$f_{eq}(N)A(N)w_-(N) = f_{eq}(N-1)A(N-1)w_+ \qquad (1)$$

from where $w_-(N)$ is found via the known probability $w_+$. Here $A(N)$ is the surface area of a cluster consisting of $N$ monomers, $f_{eq}(N)$ is the equilibrium distribution function of clusters.

Such model leads to the equation for the distribution function $f(N)$ similar to the Einstein-Smoluchowski equation in the diffusion theory [4]. In turn, the latter relates to the case of "high viscosity" [15]. In this connection, an interesting question of the cases of arbitrary and low viscosity arises which cannot be studied in the framework of the standard classical theory. This issue, as well as other, can be considered within a multivariable theory coupled with *macroscopic* kinetics.

The multivariable nucleation theory starting with the classical work of Reiss [16] has been developed in subsequent works [17-20]. It is important that not only the composition of a cluster can be the variable $x_i$ of its description (so that a multivariable theory is identified sometimes with a multicomponent one), but also another state variables (volume, density, temperature, etc. [21]); the complete set of variables $\{x_i\}$ describing the state of an embryo is determined in a specific problem. In other words, the nucleation theory is multivariable even in a single-component case. In particular, the inclusion of temperature in the theory allows investigating and evaluating the nonisothermal effect in nucleation [5, 6, 11-13].

Rationale for the use of macroscopic kinetics in nucleation phenomena is as follows. We assume that the evolution of the distribution function $f(\{x_i\},t)$ of clusters subjected by the stochastic action of the thermostat (the surrounding mother phase) is described by the Fokker-Planck equation

$$\frac{\partial f(\{x_i\},t)}{\partial t} = -\frac{\partial J_i(\{x_i\},t)}{\partial x_i} \tag{2}$$

$$J_i = -d_{ij}\frac{\partial f}{\partial x_j} + \dot{x}_i f \tag{3}$$

where $\mathbf{J}(\{x_i\},t)$ is the vector of the flux density of clusters in the space $\{x_i\}$.

The equilibrium distribution function $f_{eq}(\{x_i\})$ [19],

$$f_{eq}(\{x_i\}) = N_0\sqrt{\frac{h_{11}^{-1}\det \mathbf{H}}{(2\pi kT)^{p-1}}}\, e^{-\frac{W(\{x_i\})}{kT}} \tag{4}$$

vanishing the flux (3) gives an equation for $\dot{x}_i$:

$$\dot{x}_i = -\frac{d_{ij}}{kT}\frac{\partial W}{\partial x_j} = -\frac{d_{ij}}{kT}h_{jk}x_k \equiv -z_{ik}x_k, \quad \mathbf{Z} = \mathbf{DH}/kT \tag{5}$$

Here $W(\{x_i\})$ is the work of formation of the cluster in the state $\{x_i\}$; it can be represented in the vicinity of the saddle point $\{x_i^*\}$ as the quadratic form $W = W_* + (1/2)h_{ik}x_i x_k$; $p$ is the number of variables $x_i$; $N_0$ is the normalizing constant of the one-dimensional distribution function $f_{eq}(x_1)$. $\mathbf{D}$ is the matrix of "diffusivities", its elements $d_{ij}$ are assumed to be constant and equal to the values at the saddle point. The prefactor $N_0$ has been calculated in the framework of statistical mechanical approach [22-26] for the distribution function $f_{eq}(N)$ of clusters of $N$ molecules:

$$N_0 = \frac{\rho_{liq}}{\rho_{vap}} N_{vap} \equiv \zeta N_{vap} \tag{6}$$

where $\rho_{vap}$ and $\rho_{liq}$ are the densities of vapor and liquid, respectively; $N_{vap}$ is the total number of molecules of vapor. The factor $\zeta = \rho_{liq}/\rho_{vap}$ is the product of the replacement free energy factor, $\rho_{liq}/\rho_{vap}^{(e)}$, and the inverse supersaturation factor, $\rho_{vap}^{(e)}/\rho_{vap}$, which are shown in ref. [23] to have their origins in the inclusion of the mixing entropy in the theory (the subscript (*e*) refers to the saturated vapor). The parameter $\zeta$ also plays an important role in the theory presented.

The steady state flux of embryos, or the nucleation rate, has been calculated in a general form in ref. [19]:

$$I = N_0 \sqrt{\frac{kT}{2\pi} |h_{11}^{-1}|} |\kappa_1| e^{-\frac{W_*}{kT}} \tag{7}$$

where subscript 1 relates to the unstable variable $x_1$ (the volume $V$ of a cluster here), $\kappa_1$ is the negative eigenvalue of the matrix $\mathbf{Z}$.

As is known [27], the velocities $\dot{x}_i$ and diffusivities $d_{ij}$ in the Fokker-Planck equation are defined as follows:

$$\dot{x}_i = \lim_{\Delta t \to 0} \frac{\langle \Delta x_i \rangle_{\Delta t}}{\Delta t}, \quad d_{ij} = \lim_{\Delta t \to 0} \frac{\langle \Delta x_i \Delta x_j \rangle_{\Delta t}}{\Delta t} \tag{8}$$

where the averaging over all possible changes $\Delta x_i$ (with the corresponding probability) in the value of the quantity $x_i$ in the time $\Delta t$ is made. Thus, despite the fact that the motion of an embryo in the vicinity of the saddle point is chaotic (Brownian), the averaging procedure retains only the regular component of this motion, i.e. the quantities $\dot{x}_i$ and $d_{ij}$ as well as eq.(5) are *macroscopic*. This means that we can use macroscopic equations of motion of an embryo in the space $\{x_i\}$ for calculating the quantity $\kappa_1$ and thereby the nucleation rate. The usual hydrodynamics serves for this purpose.

The present report is devoted to the implementation of this program for the droplet nucleation in a supersaturated vapor. The paper is structured as follows. At first (Sec. 2) the thermodynamics of droplet formation is considered and the matrix $\mathbf{H}$ is obtained. The hydrodynamic flow of vapor to the droplet is the subject of Sec. 3, where the droplet growth equation is derived. The kinetics of droplet nucleation in the high-viscosity limit of the growth equation is studied in Sec. 4. Both nonisothermal and isothermal cases are considered. The thermal problem for a droplet is also studied here and the numerical estimate of the nonisothermal effect for water vapor is given. The complete growth equation (the arbitrary-viscosity case) is examined and the estimate of the viscosity effect is obtained in Sec. 5. In Sec. 6, the effect of a neutral background gas is studied. In Sec. 7, the steady state distribution function of droplets and their average temperature are calculated. The summary of results is given in Sec. 8.

## 2. Thermodynamics of droplet formation

### 2.1. Conditions of equilibrium and critical radius

The pressure $P_0$ and the temperature $T_0$ characterize the supersaturated vapor state. Hereafter the subscript 0 refers to the vapor state; the droplet state parameters are without an index. The known thermodynamic conditions of equilibrium of the critical embryo (or the embryo located at the saddle point; its parameters are denoted by asterisk) with the parent phase were confirmed in ref. [28] via the nonequilibrium treatment of the cluster formation:

$$T_* = T_0 \tag{9a}$$

$$\mu_* = \mu_0 \tag{9b}$$

$$P_* - P_0 = \frac{2\sigma}{R_*} \equiv P_L^* \tag{9c}$$

where $\mu$ is the chemical potential, $\sigma$ is the surface tension, $R$ is the droplet radius.

Eq. (9b) with account for eqs. (9a) and (9c) (asterisk is omitted),

$$\mu\left(T_0, P_0 + \frac{2\sigma}{R}\right) = \mu_0(T_0, P_0) \tag{10a}$$

yields the dependence of the saturated vapor pressure on the droplet radius, $P_0(R)$. In practice, equations for the two-phase equilibrium displacement [29] are employed for deriving this and similar dependences; in our case this is the equation

$$d\mu = d\mu_0 \tag{10b}$$

The dependence of the equilibrium surface tension on state parameters of the system "critical embryo + ambient phase" can be expressed in different ways: $\sigma(T_0, P_0)$, $\sigma(T_0, R)$, $\sigma(P_0, R)$, $\sigma(P_*, R)$, etc.; differential equations for all these dependences are given in ref. [29]. Considering eq. (10b) at $T_0 = const$ and employing the dependence $\sigma(T_0, R)$ in it, we have

$(v_0 - v)dP_0 = vd(2\sigma_{T_0}(R)/R)$, or

$$[\zeta(P_0, T_0) - 1]dP_0 = d\left(\frac{2\sigma_{T_0}(R)}{R}\right), \quad \zeta = \frac{v_0}{v} \tag{11a}$$

where $\sigma_{T_0}(R)$ is the isothermal dependence of the surface tension on radius [29]; $v$ is the volume per one molecule. We employ the approximation of incompressible liquid, so $v$ does not depend on $P_0$ and $R$. The dependence $v_0(P_0, T_0)$ is found from the equation of state of vapor, $F(P_0, v_0, T_0) = 0$. Integration of eq. (11a) gives

$$\int_{P_\infty(T_0)}^{P_0} [\zeta(P_0', T_0) - 1]dP_0' = \frac{2\sigma_{T_0}(R)}{R} \tag{11b}$$

$P_\infty(T_0)$ is the saturated vapor pressure at the planar interface ($P_\infty(T_0)$ is the coexistence curve, or the binodal, on the $(T_0, P_0)$-plane),

$$P_\infty(T_0) = P_{ref}\, e^{-\frac{q}{kT_0}} \tag{12}$$

$q$ is the heat of vaporization per one molecule (it is implied not depending on $T_0$), $P_{ref}$ is the reference pressure.

Eq. (11b) presents the desired dependence $P_0(R)$ in a general form; also this is an equation for the critical radius value, $R_*$, for given $T_0$ and $P_0$. Applying eq. (11b) to an ideal gas, $v_0(P_0, T_0) = kT_0/P_0$, $\zeta \gg 1$, we have

$$P_0(R) = P_\infty(T_0)\, e^{\frac{2v\sigma_{T_0}(R)}{kT_0 R}} \tag{13}$$

So, the form of the Kelvin equation remains for the radius-dependent surface tension also. However, differently from the classical Kelvin equation (with $\sigma = const$), when $P_0 \to \infty$ at $R \to 0$, eq. (13) gives a finite value of $P_0$ (the limiting supersaturation $\overline{P_0}$) in this case; the asymptotics

$$\sigma_{T_0}(R) = K(T_0)R \tag{14}$$

with a temperature-dependent coefficient $K$ has to be employed at $R \to 0$ [29, 30]. The difference of pressures in eq. (9c) is also finite in this limit.

It should be stressed that this is only qualitative arguments. In fact, the ideal gas approximation is hardly applicable in the region of limiting supersaturation (large value of $P_0$ and small values of $R_*$). So, eq. (13) makes sense only at sufficiently large values of $R_*$, when $\sigma = const$, and this is just the common case of its applications. The critical radius determined by this equation via the supersaturation $P_0/P_\infty$ is

$$R_* = \frac{2\upsilon\sigma}{kT_0}\left(\ln\frac{P_0}{P_\infty}\right)^{-1} \tag{15a}$$

If the ideal gas approximation is still valid at a small value of $R_*$ and the dependence $\sigma(R)$ has to be taken into account, then the critical radius is determined from the relation

$$\frac{\sigma_{T_0}(R_*)}{R_*} = \frac{kT_0}{2\upsilon}\ln\frac{P_0}{P_\infty} \tag{15b}$$

Otherwise, general eq. (11b) together with an equation of state of a real gas has to be employed for determining the critical radius.

It is seen that the coefficient $K(T_0)$ is determined by eq. (11b) also via the known limiting supersaturation $\overline{P}_0(T_0)$:

$$K(T_0) = \frac{1}{2}\int_{P_\infty(T_0)}^{\overline{P}_0(T_0)}[\zeta(P_0',T_0)-1]dP_0' \tag{16}$$

Conversely, the limiting supersaturation, or the spinodal $\overline{P}_0(T_0)$, can be calculated from the known function $K(T_0)$. We can obtain the spinodal equation by varying eq. (16) in temperature:

$$\frac{d\overline{P}_0}{dT_0} = \frac{s_{0,sp}-s}{\upsilon_{0,sp}-\upsilon} + \frac{2\upsilon}{\upsilon_{0,sp}-\upsilon}\frac{dK(T_0)}{dT_0} \tag{17}$$

where $s$ is the entropy per one molecule and the subscript $sp$ refers to the spinodal state. It can be also directly obtained from eq. (10b) applied to the spinodal state [30]:

$$(s_{0,sp}-s)dT_0 = (\upsilon_{0,sp}-\upsilon)d\overline{P}_0 - \upsilon d\left(\frac{2\sigma(T_0,R)}{R}\right) \tag{18}$$

The above consideration concerns the supersaturation with respect to the pressure $P_0$ at $T_0 = const$. The supersaturation with respect to the temperature $T_0$ at $P_0 = const$ can be considered similarly. Eq. (10b) with the dependence $\sigma(P_0,R)$ reads at $P_0 = const$ as follows:

$$\rho(T_0)\Delta s(T_0)dT_0 = -d\left(\frac{2\sigma_{P_0}(R)}{R}\right), \quad \Delta s(T_0) = s_0 - s, \quad \rho = \frac{1}{\upsilon} \tag{19}$$

from where

$$\int_{T_0}^{T_\infty(P_0)}\rho(T_0')\Delta s(T_0')dT_0' = \frac{2\sigma_{P_0}(R)}{R} \tag{20}$$

where $T_\infty(P_0)$ is the equilibrium temperature for the planar interface; $\sigma_{P_0}(R)$ is the isobaric dependence of the surface tension on radius [29].

Expanding the function $\Delta s(T_0)$ near the binodal $T_\infty(P_0)$,

$$\Delta s(T_0) = \frac{q}{T_\infty} + \frac{\Delta c_p}{T_\infty}(T_0 - T_\infty) + \ldots, \quad \Delta c_p = c_p^0 - c_p \tag{21}$$

where $c_p$ is the heat capacity per one molecule, and assuming $\rho = const$, we can calculate the integral in eq. (20a). Taking only the first term of this expansion, we have

$$T_0(R) = T_\infty - \frac{2v\sigma T_\infty}{qR} \tag{22}$$

This relation determines the equilibrium temperature of vapor for the droplet of radius $R$ in the region of large radii (near the binodal). The critical radius $R_*$ is determined by this equation via the supercooling $\Delta T = T_\infty - T_0$ of the vapor. In the region of small radii (deep supercooling), general eq. (20) have to be employed for the critical radius calculation.

It should be noted that the quantity $R$ is not the actual size of the new phase, but the size of the geometric figure bounded by the surface of tension. The difference between the locations of the physical and geometrical surfaces is negligible in comparison with $R$ at large values of $R$. However, at small values of $R$ comparable with the thickness of the surface layer, the size of the actual inhomogeneity is significantly larger than $R$. Therefore, the condition $R \to 0$ is valid and physically realistic, so that rigorous thermodynamic relations have to involve this limit [29]. The parameters $T_*$, $P_*$, and $\mu_*$ in eqs. (9a-c) relate to the bulk liquid phase. At a sufficiently small value of $R_*$, the embryo is fully inhomogeneous and seems to consist of the surface layer only; there is no real bulk phase. In this case, the bulk phase is implied as a *reference* phase; its state is uniquely defined by eqs. (9a-c). Deviations of extensive thermodynamic quantities from their bulk values are taken into account (in Gibbs' method) via the superficial quantities. So, the work of the droplet formation has the form [21]

$$W = (\mu - \mu_0)(N + \overline{N}) + (T - T_0)(S + \overline{S}) - (P - P_0)V + \sigma A \tag{23}$$

where $\overline{N}$ and $\overline{S}$ are the one-sided [28] superficial number of particles and superficial entropy of a cluster, respectively. The role of these quantities increases with decreasing $R$ and becomes essential in the case discussed.

### 2.2. Work near the saddle point

As shown above, the matrix $\mathbf{H}$ is determined by the second differential of the work at the saddle point. In ref. [21], the following expression for the second differential has been obtained:

$$d^2W = N[dTds - dPdv] - \frac{2}{9}g\sigma V^{-4/3}(dV)^2 \tag{24}$$

where $g = 3^{2/3}(4\pi)^{1/3}$; $V = N\upsilon$.

In the approximation of incompressible liquid, $d\upsilon = 0$, $\upsilon = const$, only one variable remains from $V$ and $N$; the cluster entropy is $S(V,T) = S(N\upsilon,T) = Ns(T)$, $ds = (c_V/T)dT$. Thus, eq. (24) at the saddle point is

$$\left(d^2W\right)_* = -\frac{2}{9}g\sigma V_*^{-4/3}(dV)^2 + \frac{C_V^*}{T_0}(dT)^2 \tag{25}$$

where $C_V^* = N_*c_V = c_V\rho V_*$ is the heat capacity of the critical droplet, $\rho$ is the molecular density of the liquid. Hence, the matrix $\mathbf{H}$ is

$$\mathbf{H} = \begin{pmatrix} -\frac{2}{9}g\sigma V_*^{-4/3} & 0 \\ 0 & \frac{c_V\rho V_*}{T_0} \end{pmatrix} \tag{26}$$

The matrix element $h_{TT} = N_*c_V/T_0$ determines the fluctuations of the cluster temperature $T$ around the mean (equilibrium) value $T_0$ [19]: $\langle(\Delta T)^2\rangle = kT_0/h_{TT}$. So,

$$\frac{\sqrt{\langle(\Delta T)^2\rangle}}{T_0} = \sqrt{\frac{k}{c_V}}\frac{1}{\sqrt{N_*}} \tag{27a}$$

For the water cluster ($c_V = 9.2k$) consisting of $N_* = 100$ molecules at $T_0 = 300\,°K$, eq. (27a) gives a noticeable value of the mean square fluctuation $\sqrt{\langle(\Delta T)^2\rangle} = 10\,°K$.

The mean square energy fluctuation (known from the theory of fluctuations) for an incompressible droplet is $\Delta E = \sqrt{kc_V N_*}T_0$. The mean square fluctuation of the inverse temperature is $\Delta(1/T) = \sqrt{\langle(\Delta T)^2\rangle}/T_0^2$. So the thermodynamic uncertainty relation $\Delta E\Delta(1/T) \geq k$ known from literature is minimized for an incompressible droplet:

$$\Delta E\Delta\frac{1}{T} = k \tag{27b}$$

Eq. (23) for the nucleation work and its multicomponent extension [28] were obtained for an arbitrary embryo state, when generally there is no equilibrium with the parent phase in any state parameter. Thereby they allow considering the fluctuations of different stable variables characterizing the embryo state and make the nucleation theory multivariable. The theory becomes one-dimensional, if we "forbid" the fluctuations, i.e. put $h_{TT}/kT_0 \to \infty$ in our case (the thermodynamic limit [19]). Another type of transition to the one-dimensional theory, the kinetic limit [21], is considered below.

The effect of cluster energy fluctuations on the steady state nucleation rate was studied in ref. [6]. It should be stressed that there is no need to consider the similar effect of temperature fluctuations in the presented theory, since the effect of fluctuations of all stable variables on the nucleation rate has already been taken into account by eq. (7) via the procedure of its derivation in ref. [19]. We have only to consider the kinetics of heat transfer between the droplet and the surrounding vapor in order to calculate the quantity $\kappa_1$ in this equation.

## 3. Droplet growth equation

We consider the spherically symmetric flow of vapor with the velocity $u(r,t)$ to the droplet; the vapor flux density is $j = \rho_0 u$, $\rho_0$ is the vapor particle density. Let $u_R$ and $P_R$ be the vapor velocity and pressure at the droplet surface. For determining $u_R$, we have the equation

$$\frac{dV}{dt} = -4\pi R^2 j_R \upsilon \tag{28}$$

from where

$$u_R = -\zeta \dot{R}, \quad \dot{R} \equiv dR/dt \tag{29}$$

For determining $P_R$, we employ the mass transfer equation [21]

$$\dot{N} = \frac{\pi R^2 \beta u_0}{kT_0}(P_R - P_{eq}) \tag{30}$$

where $\beta$ is the condensation coefficient (or the mass accommodation coefficient [11]), $u_0 = (8kT_0/\pi m)^{1/2}$ is the mean thermal velocity of vapor molecules; $P_{eq}(R,T)$ is the equilibrium pressure of the vapor for the droplet of radius $R$ and temperature $T$; it is given by eq. (13) with arbitrary $R$ and $T$. From the equation $dV/dt = \dot{N}\upsilon$ and eq. (30), one obtains

$$P_R = P_{eq}(R,T) + \frac{\zeta P_0}{\beta u_1}\dot{R} \tag{31}$$

where $P_0 = \rho_0 kT_0$ is the vapor pressure, $u_1 = u_0/4 = (kT_0/2\pi m)^{1/2}$.

The condition $u/c \ll 1$, where $c$ is the sound velocity, is obviously satisfied, so the vapor flow can be considered as incompressible; the continuity equation then reads as $\partial(r^2 u)/\partial r = 0$, from where $u(r,t) = C(t)/r^2$. Utilizing the boundary condition (29), we find

$$u(r,t) = -\zeta \left(\frac{R}{r}\right)^2 \dot{R} \tag{32}$$

The Navier-Stokes equation for the given case is

$$\frac{\partial u}{\partial t} + u\frac{\partial u}{\partial r} = -\frac{1}{\rho_{0m}}\frac{\partial P}{\partial R} + \nu_0\left[\Delta u - \frac{2u}{r^2}\right] \quad (33a)$$

where $\rho_{om} = \rho_0 m$ is the vapor mass density, $m$ is the molecule mass; $\nu_0$ is the vapor kinematic viscosity. It is seen that the expression in square brackets in this equation (the radial component of the Laplacian) is equal to zero for the velocity given by eq. (32). So, we have

$$\frac{1}{\rho_{0m}}\frac{\partial P}{\partial r} + \frac{\partial u}{\partial t} + u\frac{\partial u}{\partial r} = 0 \quad (33b)$$

Substituting here $u(r,t)$ from eq. (32) and then integrating eq. (33b) over $R$ from $R$ to $\infty$, we get

$$\frac{1}{\rho_{0m}}[P_\infty - P_R] - \zeta R\ddot{R} - \zeta'\dot{R}^2 = 0, \quad \zeta' \equiv \zeta^2/2 + 2\zeta \quad (34)$$

Utilizing the boundary conditions, eq. (31) and $P_\infty = P_0$, as well as the equality

$$\frac{P_0}{\beta u_1 \rho_{0m}} = \frac{2\pi}{\beta}u_1 \equiv \omega \quad (35)$$

we arrive at the following equation:

$$R\ddot{R} + \left(\frac{\zeta}{2} + 2\right)\dot{R}^2 + \omega\dot{R} = \frac{1}{\rho_m}[P_0 - P_{eq}(R,T)] \quad (36)$$

where $\rho_m = \zeta\rho_{0m}$ is the liquid mass density.

Eq. (36) is similar to the corresponding equation for the bubble dynamics in a liquid [21] except for the one significant difference: $\omega = 4\nu_{liq}/R$ for the latter, where $\nu_{liq}$ is the kinematic viscosity of the liquid. The viscosity comes in the equation from the boundary condition which is the equality of normal stresses at the bubble boundary [31]. In the case considered, the boundary conditions are quite different, so eq. (36) does not involve the vapor viscosity. We can formally define the ""effective kinematic viscosity" $\nu_{ef}$ for eq. (36) via the relation

$$\omega = \frac{4\nu_{ef}}{R_*}, \quad \nu_{ef} = \frac{1}{4}\omega R_* = \frac{\pi}{2\beta}u_1 R_* \quad (37a)$$

and the corresponding "effective dynamic viscosity" $\eta_{ef}$ as

$$\eta_{ef} = \nu_{ef}\rho_{0m} = \frac{P_0 R_*}{4\beta u_1} \quad (37b)$$

Passing from $R$ to $V$ and neglecting the term $\sim \dot{V}^2$ in the vicinity of the saddle point, we get finally:

$$\ddot{V} = \frac{3V^{1/3}}{g'\rho_m}[P_0 - P_{eq}(V,T)] - \frac{\omega}{\sqrt{g'}}\frac{\dot{V}}{V^{1/3}}, \quad g' \equiv (3/4\pi)^{2/3} \quad (38)$$

In the high-viscosity limit, we can put $\ddot{V} = 0$ [31], so the growth equation is

$$\dot{V} = \frac{gV^{2/3}}{\omega\rho_m}\left[P_0 - P_{eq}(V,T)\right] \qquad (39a)$$

In terms of $R$, it reads as

$$\dot{R} = \frac{1}{\omega\rho_m}\left[P_0 - P_{eq}(R,T)\right] \qquad (39b)$$

Eq. (39a) gives for $\dot{N}$ the following equation:

$$\dot{N} = \frac{\pi R^2 \beta u_0}{kT_0}\left[P_0 - P_{eq}\right] \qquad (39c)$$

i.e. eq. (30) with $P_0$ instead of $P_R$. Indeed, the substitution of eq. (39b) in eq. (31) yields $P_R = P_0$. The deviation $\Delta P = P_0 - P_{eq}$ of the vapor pressure from the equilibrium pressure for the droplet of a given radius and temperature is the cause of the droplet growth; it can be called the "driving force" for the droplet growth. The vapor flow in the high-viscosity limit is a consequence of the assumption $\rho_0 = const$ and the continuity equation.

## 4. Nucleation kinetics in the high-viscosity limit

### 4.1. Droplet motion equations on the $(V,T)$ plane

The general form of the motion equations in the vicinity of the saddle point is

$$\begin{cases} \dot{V} = -z_{VV}(V - V_*) - z_{VT}(T - T_0) \\ \dot{T} = -z_{TV}(V - V_*) - z_{TT}(T - T_0) \end{cases} \qquad (40)$$

The equation for $\dot{T}$ can also be written in the form [21]

$$\dot{T} = a_T \dot{V} - \lambda_{TT}(T - T_0) \qquad (41)$$

From these equations, one obtains

$$z_{TV} = a_T z_{VV}, \quad z_{TT} = \lambda_{TT} + a_T z_{VT} \qquad (42)$$

The symmetry condition $d_{VT} = d_{TV}$ for the matrix $\mathbf{D}/kT_0 = \mathbf{ZH}^{-1}$ (see eq. (5)) determines $a_T$ from the known equation for $\dot{V}$:

$$a_T = \frac{z_{VT}}{z_{VV}} \frac{h_{VV}}{h_{TT}} \qquad (43)$$

Expanding the right-hand side of eq. (39a) near the saddle point up to linear terms (the quantities $\sigma$ and $\upsilon$ are assumed constant), we find the elements $z_{VV}$ and $z_{VT}$. After simple transformations, they can be represented as follows:

$$z_{VV} = -\xi P_L^*, \quad \xi \equiv \frac{\beta u_1}{\zeta^2 P_0 R_*} = \frac{1}{4\zeta^2 \eta_{ef}} \tag{44a}$$

$$z_{VT} = 3\xi k N_* \tilde{q}, \quad \tilde{q} \equiv \frac{q - 2\upsilon\sigma/R_*}{kT_0} \tag{44b}$$

In this representation, the parameters $z_{VV}$ and $z_{VT}$ look like the corresponding parameters for the bubble nucleation [21]. From here, the parameter $a_T$ is determined:

$$a_T = \frac{kT_0 \tilde{q}}{c_V V_*} \tag{45}$$

Substituting the $a_T$ value in eq. (41) and replacing $\dot{V}$ by $\upsilon \dot{N}$, we can represent this equation in the following forms:

$$C_V^* dT = \left[ q - \frac{2\upsilon\sigma}{R_*} \right] dN - \lambda_{TT} C_V^* (T - T_0) dt \tag{46a}$$

or

$$dE = dW_\sigma + dQ, \quad dE = C_V dT, \quad dW_\sigma = -\frac{2\upsilon\sigma}{R} dN, \quad dQ = qdN - \lambda_{TT} C_V (T - T_0) dt \tag{46b}$$

The quantity $C_V^* dT = dE_*$ is the energy increment of the critical droplet. It consists of the two parts in the right-hand side of eq. (46a): (*i*) the energy increment due to the condensation of $dN$ molecules (the work of the surface increase per one molecule, $2\upsilon\sigma/R_*$, is subtracted from the heat of vaporization) and (*ii*) the energy increment due to the heat exchange with the ambient vapor. So, eq. (46a) and, accordingly, eq. (41) for $\dot{T}$ are the *energy balance equations*, as it was expected. Eq. (46b) is the representation of eq. (46a) in the form of the *first law of thermodynamics* for the droplet. The same holds for the bubble nucleation [21].

Writing for the heat flow

$$\lambda_{TT} C_V^* (T - T_0) dt = 4\pi R_*^2 \alpha (T - T_0) dt \tag{47}$$

we find

$$\lambda_{TT} = \frac{3\alpha}{c_V \rho R_*} \tag{48}$$

where $\alpha$ is the heat transfer coefficient. Finally, eq. (41) acquires the following form:

$$\dot{T} = \frac{kT_0 \tilde{q}}{c_V V_*} \dot{V} - \frac{3\alpha}{c_V \rho R_*} (T - T_0) \tag{49}$$

From the above equations, the matrix $\mathbf{Z}$ is completely determined:

$$\mathbf{Z} = \begin{pmatrix} z_{VV} & z_{VT} \\ a_T z_{VV} & \lambda_{TT} + a_T z_{VT} \end{pmatrix} = \begin{pmatrix} -\xi P_L^* & 3kN_* \xi \tilde{q} \\ -\frac{kT_0 \tilde{q}}{c_V V_*} \xi P_L^* & \lambda_{TT} + 3\xi \tilde{P} \end{pmatrix}, \quad \tilde{P} \equiv \left( \frac{k}{c_V} \tilde{q}^2 \right) \rho k T_0 \tag{50}$$

$$\det \mathbf{Z} = z_{VV}\lambda_{TT}, \quad Sp\mathbf{Z} = z_{VV} + \lambda_{TT} + 3\xi\widetilde{P} \tag{51}$$

The characteristic equation is

$$\kappa^2 - (Sp\mathbf{Z})\kappa + \det \mathbf{Z} = 0 \tag{52}$$

from where

$$\kappa_1 = \frac{1}{2}\left\{Sp\mathbf{Z} - \sqrt{(Sp\mathbf{Z})^2 - 4\det \mathbf{Z}}\right\} \tag{53}$$

The thermodynamic limit $h_{TT}/kT_0 \to \infty$ mentioned above can be formally attributed to $c_V \to \infty$; the mean-square temperature fluctuation, eq. (27a), then tends to zero. It is seen that eq. (49) in this limit has the form $\dot{T} = 0$, from where $T = const = T_0$, as it must.

### 4.2. Isothermal limit. Zeldovich-Frenkel theory

In the isothermal limit $\lambda_{TT} \gg |z_{VV}|, 3\xi\widetilde{P}$, we have $(Sp\mathbf{Z})^2 \gg |\det \mathbf{Z}|$ and $Sp\mathbf{Z} > 0$; eq. (53) yields

$$\kappa_1 = \frac{\det \mathbf{Z}}{Sp\mathbf{Z}} = z_{VV} \tag{54}$$

i.e. the theory becomes one-dimensional. Differently from the thermodynamic limit, this is the kinetic one, since it is determined by the rate of heat exchange between the droplet and vapor. As it was mentioned above, the Zeldovich-Frenkel theory corresponds to a "high-viscosity" case; in addition, it is isothermal. So, the limit considered here must give the nucleation-rate expression obtained in this theory [4]:

$$I = \sqrt{\frac{|h_{NN}|}{2\pi kT_0}} d_{NN} f_{eq}(N_*) \tag{55}$$

where

$$d_{NN} = A(N_*)w_+ = gV_*^{2/3}\rho_0 u_1 \beta \tag{56a}$$

is the coefficient of diffusion on the $N$-axis.

The elementary increment is $\Delta N = 1$, hence $d_{NN} = (\Delta N)^2/\tau = 1/\tau$, where $\tau^{-1}$ is the frequency of monomers attachment. The droplet volume $V$ is employed in the presented theory, accordingly, $d_{VV} = (\Delta V)^2/\tau = v^2/\tau$, since $\Delta V = v$ corresponds to $\Delta N = 1$. Thus,

$$d_{VV} = v^2 d_{NN} \tag{56b}$$

The coefficient $d_{VV}$ is determined by the equation $\mathbf{D} = kT_0\mathbf{Z}\mathbf{H}^{-1}$:

$$d_{VV} = kT_0 \frac{z_{VV}}{h_{VV}} \qquad (57a)$$

Substituting here the obtained above expressions for $h_{VV}$ and $z_{VV}$, as well as employing the relations $\zeta = 1/\rho_0 v$, $P_0 = \rho_0 kT_0$, $R_*^2 = g'V_*^{2/3}$, and $g^2 g' = 9$, we get

$$d_{VV} = \left[ gV_*^{2/3} \rho_0 u_1 \beta \right] v^2 = d_{NN} v^2 \qquad (57b)$$

in full accordance with eqs. (56a, b).

Multivariable nucleation-rate eq. (7) in the one-dimensional limit, $\kappa_1 = z_{VV} = d_{VV} h_{VV} / kT_0$, has the form

$$I = \sqrt{\frac{|h_{VV}|}{2\pi kT_0}} d_{VV} f_{eq}(V_*) \qquad (58)$$

In view of the relations $h_{VV} = h_{NN}/v^2$, $f_{eq}(V) = f_{eq}(N)/v$ and eq. (56b), eq. (58) is identical to eq. (55), i.e. the nucleation rate is invariant with respect to the choice of the embryo description variable, as it must. So, the high-viscosity isothermal limit yields the classical Zeldovich-Frenkel nucleation-rate equation, as it was expected.

### 4.3. Nonisothermal nucleation

The energy balance eq. (46a) can be written for an arbitrary-size droplet also (if the temperature difference $T - T_0$ is not too large). So, the system of eqs. (36) and (49) describes the droplet growth and enables to find the dependence $R(t)$, if the heat transfer coefficient $\alpha$ is known. The calculation of this coefficient is a separate problem. Way of describing the heat exchange between the droplet and surrounding vapor depends on the ratio $R/l$, where $l$ is the mean free path of a vapor molecule; the droplet radius $R$ is the characteristic size in the considered thermal problem. Below three cases for this ratio are considered and numerical estimates of $\alpha$ and $\lambda_{TT}$ for water vapor at $T_0 = 300\,°K$ are made. The physical properties of water and vapor together with some calculated parameters are given in the Table. The critical radius value $R_* = 10^{-7}\,cm$ corresponding to the pressure supersaturation $P_0/P_\infty = e$ is employed in all calculations. This value is of the order of the range of intermolecular forces and therefore assumed to be the lowest one, when we can use (approximately) the macroscopic (not depending on $R$) surface tension; the dependence $\sigma(R_*)$ has to be taken into account for $R_* < 10^{-7}\,cm$. Of course, precise calculations have to take into account the dependence $\sigma(R_*)$ at $R_* \geq 10^{-7}\,cm$ also, since the nucleation rate is very sensitive to this quantity via the exponential function; however, we are interested only in the qualitative picture here. At the same time, such droplet contains about a hundred molecules and can be assumed as macroscopic. Also, the ideal gas approximation is applicable until the mean intermolecular spacing $a$ in the vapor becomes equal to about $10^{-7}\,cm$.

The order of $l$ for gases under normal conditions ($P_0 = 1\,bar$) is $10^{-5}\,cm$. In our case, the vapor pressure is $P_0 = eP_\infty = 0.1\,bar$. The dependence of $l$ on pressure is $l \sim 1/P_0$, hence $l \sim 10^{-4}\,cm$.

### 4.3.1. Hydrodynamic limit $l/R \ll 1$

The vapor is a continuous medium for the droplet in this case. The approximation of uniform temperature inside the droplet is employed here, so we consider only the external heat problem. Also, we use the steady approximation $\partial T/\partial t = 0$; its validity is assessed below. So, we have the equation

$$\mathbf{u}\nabla T = \chi_0 \Delta T \tag{59}$$

with boundary conditions $T(R) = T$ and $T(\infty) = T_0$. The velocity $u$ is given by eq. (32) with some mean value of $\dot{R}$; $\chi_0 = \theta_0/c_p^0 \rho_0$ is the vapor thermal diffusivity. The solution of this equation is

$$T(r) = T_0 - \frac{T-T_0}{e^{\gamma/R}-1} + \frac{T-T_0}{e^{\gamma/R}-1} e^{\gamma/r}, \quad \gamma \equiv \zeta R^2 \dot{R}/\chi_0 \tag{60}$$

The heat flux is

$$j = -\theta_0 \frac{dT}{dr}\bigg|_{r=R} = \theta_0 \frac{\gamma}{R^2} \frac{e^{\gamma/R}}{e^{\gamma/R}-1}(T-T_0) = \alpha(T-T_0) \tag{61}$$

from where

$$\alpha = \frac{\theta_0}{R} f\left(\frac{\gamma}{R}\right), \quad f\left(\frac{\gamma}{R}\right) \equiv \frac{\gamma}{R} \frac{e^{\gamma/R}}{e^{\gamma/R}-1} \tag{62}$$

In the limit $\gamma \to 0$, when the convective term in eq. (59) is neglected, we have the steady solution of the thermal problem in an immobile gas,

$$T(r) = T_0 + (T-T_0)\frac{R}{r}, \quad j = \frac{\theta_0}{R}(T-T_0) \tag{63}$$

from where $\alpha = \theta_0/R$; $f(\gamma/R) \to 1$ in this case.

The condition of applicability of the steady approximation is the characteristic time of heat propagation, $R^2/\chi_0$, is much less than the characteristic time $\lambda_{TT}^{-1}$ of the droplet temperature change: $\lambda_{TT} R^2/\chi_0 \ll 1$. Utilizing eqs. (48) and (62), we get

$$\frac{\lambda_{TT} R^2}{\chi_0} \sim \frac{\rho_0}{\rho} = \zeta^{-1} \ll 1 \tag{64}$$

as it must.

For numerical estimates, we take $R = 10^{-2}$ cm obeying the condition $l/R \ll 1$. Eq. (39b) is employed for estimating $\dot{R}$. The quantity $P_{eq}$ for the given $R$ and $T = T_0$ is equal to $P_\infty$ with great accuracy. So, we have $\dot{R} = 2.6 \times 10^{-2}$ cm/s, $\chi_0 = 1.4$ cm$^2$/s, $\gamma/R = 2.76$, $f(\gamma/R) = 3$. As is seen, the effect of vapor movement on the heat exchange becomes noticeable at sufficiently large values of $R$. Finally, $\alpha = 6 \times 10^5$ erg/cm$^2$s °K and $\lambda_{TT} = 4.2$ s$^{-1}$.

### 4.3.2. Case of a slightly rarefied gas, $l/R < 1$

In the case, when the ratio $l/R$ is not very small, there exists the *temperature jump* $\Delta T = T_0 - T$ on the droplet surface which is proportional to the temperature gradient in the vapor [32]:

$$\Delta T = \varsigma \frac{\partial T}{\partial r} \tag{65}$$

where the temperature jump coefficient $\varsigma$ is of the order of the mean free path $l$. Comparing this equation with the heat flux equation $j = -\theta_0 \partial T/\partial r$, we find

$$j = \frac{\theta_0}{\varsigma}(T - T_0) \tag{66}$$

from where

$$\alpha = \frac{\theta_0}{\varsigma} \tag{67}$$

Taking $R = 10^{-3}$ cm and $\varsigma \sim l = 10^{-4}$ cm for numerical estimates, we get $\alpha \sim 10^7$ erg/cm$^2$s °K and $\lambda_{TT} \sim 10^3$ s$^{-1}$.

### 4.3.3. Case of a highly rarefied gas, $l/R > 1$

Just this case covers nucleation phenomena; for $R_* = 10^{-7}$ cm, we have $l/R \sim 10^3 \gg 1$. Compared with the previous case, there is no the temperature gradient in the vapor now; we have the droplet with temperature $T$ and the vapor with temperature $T_0$. The heat transfer coefficient is calculated in the kinetic theory of gases and has the form [11, 32]

$$\alpha = \beta_\varepsilon (1-\beta)\rho_0 u_1 (c_V^0 + k/2) \tag{68}$$

where $\beta_\varepsilon$ is the thermal accommodation coefficient of the vapor molecules. The values of $\beta$ and $\beta_\varepsilon$ reported in literature [33] for water vapor are $\beta = 0.04$ and $\beta_\varepsilon = 1$. The calculated values of $\alpha$ and $\lambda_{TT}$ are $\alpha = 1.7 \times 10^7 \ erg/cm^2 s\ °K$ and $\lambda_{TT} = 1.2 \times 10^7 \ s^{-1}$.

If the hydrodynamic expression $\alpha = \theta_0/R$ used in eq. (48) for all values of $R$, then $\lambda_{TT}$ would be proportional to $1/R^2$ and equal to $1.4 \times 10^{10} \ s^{-1}$ at $R_* = 10^{-7} \ cm$ which is a strongly overestimated value. Thus, both the quantities $\alpha$ and $\lambda_{TT}$ increase with decreasing $R$, however the law of increasing in $\lambda_{TT}$ is not proportional to $1/R^2$ because of the different regimes of heat exchange at different $R$.

### 4.3.4. Nonisothermal effect in nucleation

Nonisothermal effect is estimated by the ratio of the actual nucleation rate $I$ to the isothermal one $I_{iso}$. From eq. (7), we have

$$\psi \equiv \frac{I}{I_{iso}} = \frac{\kappa_1}{z_{VV}} \tag{69}$$

with $\kappa_1$ given by eq. (53). In addition to the above value of $\lambda_{TT}$, the calculated values of $z_{VV}$ and $3\xi\widetilde{P}$ entering in $Sp\mathbf{Z}$ are $z_{VV} = -4.3 \times 10^5 \ s^{-1}$ and $3\xi\widetilde{P} = 3.8 \times 10^7 \ s^{-1}$. Eq. (69) yields $\psi = 1/4$, i.e. the nonisothermal effect actually exists for water vapor at the given conditions.

It is of interest to examine the dependence of $\psi$ on the condensation coefficient $\beta$ considering the latter as a free parameter of the theory. As follows from above equations, all the parameters entering in $Sp\mathbf{Z}$ contain $\beta$; $z_{VV}$, $3\xi\widetilde{P} \sim \beta$, whereas $\lambda_{TT} \sim (1-\beta)$. The dependence $\psi(\beta) = \kappa_1(\beta)/z_{VV}(\beta)$ is shown in Fig. 1. At $\beta \to 0$, we have the isothermal nucleation condition $\lambda_{TT} \gg |z_{VV}|, 3\xi\widetilde{P}$, so that $\psi(\beta) \to 1$. In the opposite case $\beta \to 1$, we have $\lambda_{TT} \ll |z_{VV}|, 3\xi\widetilde{P}$; in addition, $|z_{VV}| \ll 3\xi\widetilde{P}$ at the given conditions. Eq. (53) gives the following asymptotics:

$$\kappa_1 = -\frac{P_L^*}{3\widetilde{P}} \lambda_{TT} \tag{70}$$

i.e. nucleation is limited by the heat transfer. So, the nonisothermal effect is strongly pronounced in the case of $\beta$ closed to unity, since the number of reflected molecules is small and the heat is almost not removed from the drop.

Summarizing, nucleation vanishes in both the limiting cases $\beta \to 0$ and $\beta \to 1$ due to $\kappa_1 = z_{VV} \sim \beta \to 0$ and $\kappa_1 \sim \lambda_{TT} \sim (1-\beta) \to 0$, respectively.

## 5. Arbitrary-viscosity case

The variable $U = \dot{V}$ is used for studying the viscosity effect in nucleation; the theory presented in this Section is quite similar to the arbitrary-viscosity theory of bubble nucleation [31], so it is expounded briefly here. Equations of the droplet motion in the space $\{V,T,U\}$ are

$$\begin{cases} \dot{V} = -z_{VU}U = U \\ \dot{T} = -z_{TV}(V-V_*) - z_{TT}(T-T_0) - z_{TU}U \\ \dot{U} = -z_{UV}(V-V_*) - z_{UT}(T-T_0) - z_{UU}U \end{cases} \quad (71)$$

Expanding the right-hand side of eq.(38) for $\dot{U} = \ddot{V}$ near the saddle point up to linear terms we find the coefficients in the eq. (71) for $\dot{U}$:

$$z_{UV} = -\zeta^{-1}\frac{8\pi}{3}\frac{\sigma}{\rho_m V_*}, \quad z_{UT} = \frac{4\pi R_*}{\rho_m}\frac{P_0}{T_0}\tilde{q}, \quad z_{UU} = \frac{\omega}{R_*} = \frac{4\nu_{ef}}{R_*^2} \quad (72)$$

The matrix $\mathbf{H}$, eq. (26), becomes a $3\times 3$ matrix now; it is complemented by the element $h_{UU} = M$ which is the droplet "mass". It should be emphasized that this is not the actual mass; this is the "mass" with respect to the "velocity" $U$, so that the "kinetic energy" $MU^2/2$ added to the work of the droplet formation has the energy dimensionality. Composing the matrix $\mathbf{D}/kT_0 = \mathbf{ZH}^{-1}$ and applying to it the symmetry conditions $d_{VT} = d_{TV}$, $d_{VU} = -d_{UV}$, and $d_{TU} = -d_{UT}$, we get

$$\begin{cases} z_{TV}h_{VV}^{-1} = 0 \\ z_{UV}h_{VV}^{-1} = M^{-1} \\ z_{UT}h_{TT}^{-1} = -z_{TU}M^{-1} \end{cases} \quad (73)$$

Denoting $z_{TU} \equiv -a_T$ and $z_{TT} \equiv \lambda_{TT}$, we find from eqs. (73) $z_{TV} = 0$,

$$M = \frac{h_{VV}}{z_{UV}} = \zeta\frac{\rho_m}{4\pi R_*} \quad (74a)$$

$$a_T = z_{UT}\frac{M}{h_{TT}} = \frac{kT_0\tilde{q}}{c_V V_*} \quad (74b)$$

which is eq. (45).

So, the matrix $\mathbf{Z}$ is completely determined:

$$\mathbf{Z} = \begin{pmatrix} 0 & 0 & -1 \\ 0 & \lambda_{TT} & -a_T \\ z_{UV} & z_{UT} & z_{UU} \end{pmatrix} \tag{75}$$

$$\det \mathbf{Z} = z_{UV}\lambda_{TT}, \quad z_{UV} = z_{VV}z_{UU}, \quad Sp\mathbf{Z} = \lambda_{TT} + z_{UU} \tag{76}$$

The characteristic equation is

$$\kappa^3 - (Sp\mathbf{Z})\kappa^2 + B\kappa - \det \mathbf{Z} = 0 \tag{77}$$

where $B = \lambda_{TT} z_{UU} + a_T z_{UT} + z_{UV}$. In the isothermal limit $\lambda_{TT} \to \infty$, this equation reduces to the following one:

$$\kappa^2 - z_{UU}\kappa + z_{UV} = 0 \tag{78}$$

The solution of this equation,

$$\kappa_1 = \frac{1}{2}\left\{z_{UU} - \sqrt{z_{UU}^2 + 4|z_{UV}|}\right\} \tag{79}$$

in the low-viscosity limit $z_{UU}^2 \ll |z_{UV}|$, or $z_{UU} \ll |z_{VV}|$, acquires the form

$$\kappa_1 = -\sqrt{|z_{UV}|} \tag{80}$$

Substitution of eq. (80) in eq. (7) yields the low-viscosity nucleation rate

$$I = \sqrt{\frac{kT_0}{2\pi M}} f_{eq}(V_*) = \overline{U} f_{eq}(V_*) \tag{81}$$

where $\overline{U} = \sqrt{kT_0/2\pi M}$ is the mean thermal "velocity" of the droplet with the "mass" $M$ moving towards the barrier (the velocity and the mass in the space with the coordinate $V$). Thus, we have the case of Eyring's theory of the activated complex [34]; eq. (81) corresponds to the inertial motion of droplets without friction over the barrier.

So, formally the theory yields all the results of the arbitrary-viscosity theory [31] for droplets also. The final and most significant step is to estimate the ratio $z_{UU}/|z_{VV}|$ and determine the conditions, when the viscosity effect is important (similarly to the nonisothermal effect, the viscosity effect can be defined as the discrepancy between the actual and high-viscosity nucleation rates). With the use of eqs. (44a) and (72), one obtains

$$\frac{z_{UU}}{|z_{VV}|} = \frac{2\pi\zeta}{\beta^2}\left(\ln\frac{P_0}{P_\infty}\right)^{-1} \gg 1 \tag{82}$$

due to $\zeta \gg 1$ and $\beta < 1$. Thus, the low-viscosity and even intermediate cases cannot be realized in the region of applicability of the ideal gas approximation; because of the large difference in the specific volumes $\upsilon$ and $\upsilon_0$, condition (82) permits the high-viscosity case only. If the viscosity effect exists for the droplet nucleation, it can be only beyond the ideal gas approximation. The calculation of nucleation parameters, as well as ratio (82), in this case has to be based on the equation of state of a real gas; Kelvin's eq. (13) underlying the presented calculations has to be replaced by general eq. (11b) for $P_0(R)$ and $P_{eq}(R,T)$.

## 6. Effect of a neutral background gas

### 6.1. Droplet growth equation

We consider here the case, when the amount of an inert gas is much greater than the amount of vapor. In this case, the droplet grows due to the diffusion of vapor molecules, similarly to the growth of a precipitation in a dilute solution. Such situation occurs in the air, where the partial pressure of the saturated water vapor at $T_0 = 300\,°K$ is $3{,}57 \times 10^{-2}\,bar$. So, if we neglect the solubility of the air in water (which is of the order of $10^{-5}$) and therefore consider the one-component nucleation problem, this model describes the growth of fog droplets in the air.

We find the vapor density distribution $\rho_0(r)$ around the growing droplet from the steady equation $\Delta \rho_0(r) = 0$ with boundary conditions $\rho_0(\infty) = \rho_0$ and $\rho_0(R) = \rho_R$:

$$\rho_0(r) = \rho_0 - (\rho_0 - \rho_R)\frac{R}{r} \tag{83a}$$

from where the steady flux density is

$$j_R = -D\frac{d\rho_0}{dr}\bigg|_{r=R} = -D\frac{\rho_0 - \rho_R}{R} \tag{83b}$$

$D$ is the vapor diffusion coefficient.

The droplet growth equation is

$$\dot{R} = -j_R \upsilon = D\upsilon\frac{\rho_0 - \rho_R}{R} \tag{84}$$

Substituting here $\rho_R = P_R/kT_0$ and utilizing eq. (31) for $P_R$, we get

$$\dot{R} = \frac{\tilde{u}\upsilon}{kT_0}[P_0 - P_{eq}(R,T)], \quad \tilde{u} \equiv \left(\frac{R}{D} + \frac{1}{\beta u_1}\right)^{-1} \tag{85}$$

Noting that $\beta \upsilon u_1/kT_0 = 1/\omega\rho_m$, we have the following two limiting cases:

$$\dot{R} = \begin{cases} \dfrac{1}{\omega\rho_m}[P_0 - P_{eq}(R,T)], & (R/D) \ll (\beta u_1)^{-1} \\ \dfrac{D\upsilon}{RkT_0}[P_0 - P_{eq}(R,T)], & (R/D) \gg (\beta u_1)^{-1} \end{cases} \tag{86}$$

The first of these equations (the fast-diffusion limit, $D \to \infty$) coincides with eq. (39b); it describes the interface-controlled growth. The second equation (the slow-diffusion limit, $D \to 0$) corresponds to the diffusion-limited growth.

The numerical estimates are as follows. $D \sim u_0 l \sim 1 \; cm^2/s$, $\dot{R} = |z_{VV}| \Delta R \sim 10^{-3} \; cm/s$ for $\Delta R = 0.1 R_* = 10^{-8} \; cm$; $(R_*/D) \sim 10^{-7} \; s/cm$, $(\beta u_1)^{-1} \sim 10^{-3} \; s/cm$. Thus, the condition $(R_*/D) \ll (\beta u_1)^{-1}$ is satisfied and the first of eqs. (86) holds. The steady approximation applicability condition is the characteristic time of diffusion, $R_*^2/D$, much less than the characteristic time $R_*/\dot{R}$ of the droplet radius change: $(D/R_*\dot{R}) \gg 1$. Substituting here the above values, we get $(D/R_*\dot{R}) \sim 10^{10}$.

The influence of an inert gas on nucleation is determined by its participation in the heat exchange with the droplet and changing the nucleation work [14]. Below both these effects are considered.

### 6.2. Effect of a carrier gas on nucleation work

Derivation of an equation for the work $W$ of droplet formation in the presence of an inert carrier gas made according to the procedure of ref. [21] shows that the change in eq. (23) concerns only the $PV$-term:

$$W = (\mu - \mu_0)(N + \overline{N}) + (T - T_0)(S + \overline{S}) - (P - P_{tot})V + \sigma A, \quad P_{tot} \equiv P_0 + P_g \tag{87}$$

where $P_g$ is the carrier gas pressure. Eq. (9c) is replaced by the following one:

$$P_* - P_{tot} = \frac{2\sigma}{R_*} \tag{88}$$

so that eq. (10a) now reads as

$$\mu\left(T_0, P_{tot} + \frac{2\sigma}{R_*}\right) = \mu_0(T_0, P_0) \tag{89}$$

In order to find the dependence of the critical radius on the inert gas pressure, $R_*(P_g)$, we consider eq. (10b) at $T_0 = const$ and $P_0 = const$:

$$d\left(P_g + \frac{2\sigma}{R_*}\right) = 0 \tag{90a}$$

from where

$$P_g + \frac{2\sigma}{R_*} = C \tag{90b}$$

This equation is not so simple as it seems at first sight, since the surface tension generally depends on both $P_g$ and $R_*$. From different types of dependences of the surface tension listed above, the dependences $\sigma(P_{tot}, R_*)$, $\sigma(T_0, R_*)$, $\sigma(T_0, P_{tot})$ are relevant here; we choose the latter as the most convenient one, since at the given conditions it converts to the isothermal dependence on the carrier gas pressure, $\sigma_{T_0}(P_g)$. From equations given in ref. [29], the approximate dependence $\sigma_{T_0}(P_g)$ can be derived as follows:

$$\sigma_{T_0}(P_g) = \sigma_0 - K_p P_g \tag{91}$$

where $K_p$ is some coefficient, $\sigma_0$ is the surface tension at $P_g = 0$. So, the surface tension decreases with increasing $P_g$ [29].

The integration constant corresponds to the critical radius $R_*^0$ for $P_g = 0$:

$$C = \frac{2\sigma_0}{R_*^0} \equiv P_g^0 \quad, \quad R_*^0 \equiv R_*(P_0, T_0) \tag{92}$$

i.e. $R_*^0$ is determined by Kelvin's eq. (13) for the given vapor state $(P_0, T_0)$. Thus, the desired dependence $R_*(P_g)$ is

$$R_*(P_g) = R_*^0 \left(1 - \frac{K_p}{\sigma_0} P_g\right)\left(1 - \frac{P_g}{P_g^0}\right)^{-1} \tag{93a}$$

If $\sigma = const = \sigma_0$ is employed in eq. (90b), then

$$R_*(P_g) = R_*^0 \left(1 - \frac{P_g}{P_g^0}\right)^{-1} \tag{93b}$$

The nucleation work is determined by Gibbs' equation $W_* = (1/3)\sigma A_* = (4\pi/3)\sigma R_*^2(P_g)$.

Eqs. (93a, b) show that the critical radius increases with increasing $P_g$. From the condition that the function (93a) is ascending (and thereby has a physical meaning), it follows that the coefficient $K_p$ obeys the inequality

$$\frac{2K_p}{R_*^0} < 1 \tag{94}$$

Accordingly, the nucleation work also increases and the nucleation rate exponentially decreases [14].

For water vapor at the Table conditions, we have $P_g^0 = 1.4 \times 10^9 \, dyn/cm^2 \sim 10^3 \, bar$. So, the considered effect of the inert gas pressure on the nucleation rate is small in the range of $P_g = 0 \div 10\,bar$ employed here. On the other hand, the ideal gas approximation validity must be verified for those $P_g$-values, when this effect becomes noticeable.

### 6.3. Effect of a carrier gas on heat exchange

The heat transfer coefficient for the two-component mixture "vapor + inert gas" is [11, 32]

$$\alpha = \beta_\varepsilon(1-\beta)\rho_0 u_1(c_V^0 + k/2) + \beta_{\varepsilon,g}\rho_g u_{1,g}(c_{V,g}^0 + k/2) \tag{95}$$

where the inert gas parameters denoted by the subscript $g$ have the same meaning as the vapor parameters in the first addend; in particular, $\beta_{\varepsilon,g}$ is the thermal accommodation coefficient of gas molecules. So, the $\lambda_{TT}$ parameter increases with respect to the above value for the pure vapor. This is only the change in the matrix $\mathbf{Z}$, since eqs. (39b) and (39a) hold in the considered case, as before, and the change in the critical radius is negligible. So, the nonisothermal effect considered above has to diminish in the presence of a neutral background gas.

Argon is employed here as an inert gas for numerical estimates of the mentioned effect. Fig. 2 shows the nonisothermal effect as a function of the argon pressure, $\psi(P_g)$, for different values of the condensation coefficient $\beta$. Since eq. (39b) is valid in both the limiting cases of $P_g = 0$ and $P_g \gg P_0$, the range of $P_g$ in Fig.2 includes these limits. As it was expected, the deviation of the nucleation rate from the isothermal one decreases with increasing $P_g$ [11]. Differently from the pure vapor case, nucleation does not vanish in the limit $\beta \to 1$; the heat transfer is performed by inert gas molecules.

### 7. Steady state distribution function of droplets

For studying the distribution function of droplets, we use the dimensionless variables $x = (V - V_*)/V_*$ and $y = (T - T_0)/T_0$. The matrices $\mathbf{H}$ and $\mathbf{Z}$ change as follows:

$$h_{xx} = h_{VV}V_*^2, \quad h_{yy} = h_{TT}T_0^2 \ ; \quad z_{xx} = z_{VV}, \quad z_{yy} = z_{TT}, \quad z_{xy} = z_{VT}\frac{T_0}{V_*}, \quad z_{yx} = z_{TV}\frac{V_*}{T_0} \tag{96}$$

The multivariable steady state distribution function $f_s(\mathbf{r})$ has been derived in ref. [19]:

$$f_s(\mathbf{r}) = \frac{1}{2}f_{eq}(\mathbf{r})erfc(\phi(\mathbf{r})), \quad \phi(\mathbf{r}) \equiv -\frac{\mathbf{eHr}}{\sqrt{2kT_0|\kappa_b(\mathbf{e})|}}, \quad \mathbf{r} = \begin{pmatrix} x \\ y \end{pmatrix} \tag{97}$$

where $\mathbf{e} = (\cos\theta, \sin\theta)$ is the unit vector of the droplets flux direction on the $(x, y)$-plane;

$$\kappa_b(\mathbf{e}) = \frac{h_{xx} + h_{yy} \tan^2 \theta}{1 + \tan^2 \theta} \tag{98}$$

is the barrier curvature (or the curvature of the normal section of the saddle surface) in this direction. So, the function $\phi(\mathbf{r})$ acquires the form

$$\phi(x, y) = -\frac{h_{xx} x + (h_{yy} \tan \theta) y}{\sqrt{2kT_0 |h_{xx} + h_{yy} \tan^2 \theta|}} \tag{99}$$

Equation for $\tan \theta$ is [19]

$$\tan \theta = \frac{1}{2 z_{xy}} \left\{ (z_{yy} - z_{xx}) - \sqrt{(z_{yy} - z_{xx})^2 + 4 z_{xy} z_{yx}} \right\} \tag{100}$$

It should be recalled that eq. (97) for $f_s(\mathbf{r})$ has been obtained in the vicinity of the saddle point; the equilibrium distribution function

$$f_{eq}(x, y) = \zeta \rho_0 N_* \sqrt{\frac{h_{yy}}{2\pi kT_0}} e^{\frac{1}{kT_0} \left\{ -W_* + \frac{1}{2} \left( |h_{xx}| x^2 - h_{yy} y^2 \right) \right\}} \tag{101}$$

contains the quadratic expansion of the work $W(x, y)$ in this region. We take $\Delta = 0.3$ as the characteristic halfwidth of this region in $x$; we cannot extend the function $f_s(x, y)$ far beyond this value. Our main interest is the temperature distribution of droplets. In accordance with what was said just now, we define the functions

$$f_\Omega(y) = \frac{\int_\Omega f_s(x, y) dx}{\int_\Omega dx \int_{-\infty}^{+\infty} dy f_s(x, y)}, \quad f_{eq,\Omega}(y) = \frac{\int_\Omega f_{eq}(x, y) dx}{\int_\Omega dx \int_{-\infty}^{+\infty} dy f_{eq}(x, y)} \tag{102}$$

which are the temperature distributions of droplets in some region $\Omega$ in $x$; $f_\Omega(y) dy$ is the fraction of droplets with the temperatures in the interval $[y, y + dy]$ in the region $\Omega$. Hence, the mean temperature of droplets in this region is

$$\overline{T}_\Omega = T_0 (\overline{y}_\Omega + 1), \quad \overline{y}_\Omega = \int_{-\infty}^{+\infty} y f_\Omega(y) dy \tag{103}$$

It depends on the $\Omega$-region size, which is denoted by the corresponding subscript.

The functions $f_s(x, y)$ and $f_{eq}(x, y)$ for the pure water vapor at the Table conditions are given in Fig. 3. The functions $f_\Omega(y)$ and $f_{eq,\Omega}(y)$ for overcritical droplets in the region $\Omega = [0, \Delta]$ are shown in Fig. 4. It is seen that the steady state function $f_\Omega(y)$ is shifted relatively the equilibrium function $f_{eq,\Omega}(y)$ to higher temperatures. This means that droplets are overheated in average [13].

The average temperature $\overline{T}_\Omega(\beta)$ of droplets for different regions as a function of the condensation coefficient (considered as a free parameter of the theory) is shown in Fig. 5. The overheat vanishes at $\beta \to 0$ (the isothermal limit); the functions $f_\Omega(y)$ and $f_{eq,\Omega}(y)$ in Fig. 4 would coincide in this limit and give $\overline{T}_\Omega = T_0$. It is interesting that the sharp dependence $\overline{T}_\Omega(\beta)$ takes place for $\beta$-values less than about $0.1$, whereas for $\beta > 0.1$ the average temperature is almost constant. The same takes place for the dependence $\psi(\beta)$ in Fig. 1: it is sharp for $\beta < 0.1$ and then slowly drops to zero. In Fig. 6, the dependence $\theta(\beta)$ of the angle between the droplets flux direction and the $x$-axes (which is the steepest descent direction of the saddle surface $W(x, y)$) on the condensation coefficient is presented; $\tan\theta = \theta$ at these small $\theta$-values, so that the dependence of $\tan\theta$ on $\beta$ is similar. The comparison of Fig.'s 5 and 6 shows the similarity of both the dependences, so we can assume the proportionality

$$\overline{y}_\Omega \sim \tan\theta \tag{104}$$

This assumption can be confirmed, if we expand the function $erfc(\phi(\mathbf{r}))$ near the saddle point, $erfc(\phi(\mathbf{r})) = 1 - (2/\sqrt{\pi})\phi(x, y)$, neglect the term $\sim \tan^2\theta$ in eq. (91) and then evaluate the integrals in eqs. (102) and (103).

One more conclusion from Fig. 5 is the mean temperature is greater for larger droplets. Finally, Fig. 7 shows the dependence of the average temperature on the inert gas (argon) pressure. It is seen that the overheating decreases with increasing pressure, as it must.

It should be emphasized that the overheating phenomenon relates to the steady state (*non-equilibrium*) distribution of droplets, so it does not contradict to the thermodynamic condition of equilibrium, eq. (9a), generalized in accordance with Einstein's theory of fluctuations as

$$T_* = T_0 \pm \Delta T \tag{105}$$

where $\Delta T$ is the droplet temperature fluctuations in the equilibrium state. In other words, even in the equilibrium state there are *equal* fractions of overheated and supercooled droplets, so that the average temperature of droplets in any region is equal to the vapor temperature $T_0$, differently from the steady state. So, there is no need to revise the fundamentals of thermodynamics for explaining this and other phenomena, as discussed in ref. [28]. As is seen from the foregoing, all the considered nucleation effects are naturally explained within the classical and statistical thermodynamics.

## 8. Summary and conclusions

Macroscopic-kinetics approach in the theory of droplet nucleation is based on the droplet growth equation. This equation is derived with the use of the Navier-Stokes equation of the vapor movement to the droplet; it contains both $\dot{V}$ and $\ddot{V}$, i.e. this is a second-order equation similar to the equation for the bubble dynamics in a liquid. Differently from the latter, it does not involve the vapor viscosity, so that the effective viscosity is defined. Both the high- and arbitrary-viscosity cases of the growth equation are considered. The criterion of the viscosity effect significance shows that this effect is absent within the ideal gas approximation. The question whether or not it exists beyond this approximation requires a separate study. Thus, the theory of droplet nucleation in an ideal vapor is an essentially high-viscosity theory; this is a two-dimensional $(V,T)$-theory. It goes into the one-dimensional $(V)$-theory in either thermodynamic ($h_{TT}/kT_0 \to \infty$) or kinetic ($\lambda_{TT} \to \infty$) limits.

The droplet motion equations on the $(V,T)$-plane – equations for $\dot{V}$ and $\dot{T}$ – in the near-critical region are written using the previously developed algorithm [21] based on the linked-fluxes analysis [6] and Onsager's reciprocal relations. The equation for $\dot{T}$ obtained in this way is an energy balance equation, or the first law of thermodynamics, for the droplet. The two-dimensional $(V,T)$-theory is shown to become the one-dimensional Zeldovich-Frenkel theory in the isothermal limit $\lambda_{TT} \to \infty$. For the estimate of nonisothermal effect, the problem of heat transfer between the droplet and the surrounding vapor is considered and the expression for the heat transfer coefficient in the model of a highly rarefied gas is employed. Numerical estimates for the water vapor show the presence of the nonisothermal effect. This effect is shown to be strongly pronounced for the condensation coefficient $\beta$ values closed to unity. An inert background gas reduces the nonisothermal effect, so that the droplet nucleation occurs even in the case of $\beta \to 1$; the heat is removed from the droplet by gas molecules. However, this is true only for a certain range of the gas pressure $P_g$. At sufficiently high $P_g$-values, the competing action of the inert carrier gas – to increase the nucleation work – manifests itself, so that the resulting nucleation rate decreases [14].

The steady state distribution function of droplets calculated in the near-critical region shows their overheating in average. This phenomenon, as well as the deflection of the droplets flux from the steepest descent direction, is also a manifestation of the nonisothermal effect. The superheat degree is higher for larger droplets; it decreases with increasing pressure of the background inert gas.

Thus, the multivariable theory coupled with macroscopic kinetics yields the most complete and comprehensive picture of the droplet nucleation; it allows us to explore all the possible nucleation effects.

| parameters | $v$, $cm^3$ | $a$, $cm$ | $\theta_0$, $erg/cm\ s\ K$ | $c_V$, $erg/K$ | $P_\infty$, $dyn/cm^2$ | $q/kT_0$ | $\sigma$, $erg/cm^2$ |
|---|---|---|---|---|---|---|---|
| water | $3\times10^{-23}$ | $3\times10^{-8}$ | $6.1\times10^4$ | $9.2k$ | | 17.8 | 72 |
| vapor | $4.27\times10^{-19}$ | $7.5\times10^{-7}$ | $1.85\times10^3$ | $3.2k$ | $3.57\times10^4$ | | |

| $\zeta$ | $\beta$ | $\dfrac{2v\sigma}{kT_0 R_*}$ | $|z_{VV}|$, $s^{-1}$ | $\lambda_{TT}$, $s^{-1}$ | $z_{UU}$, $s^{-1}$ | $u_1$, $cm/s$ | $\dfrac{|h_{xx}|}{kT_0}$ | $\dfrac{h_{yy}}{kT_0}$ |
|---|---|---|---|---|---|---|---|---|
| $1.4\times10^4$ | 0.04 | 1 | $4.3\times10^5$ | $1.2\times10^7$ | $2.3\times10^{13}$ | $1.5\times10^4$ | 48.6 | 1283 |

Table. Physical properties of water and vapor at the temperature $T_0 = 300\ °K$ together with calculated nucleation parameters for the droplet of critical radius $R_* = 10^{-7}\ cm$ corresponding to the supersaturation in pressure $P_0/P_\infty = e$.

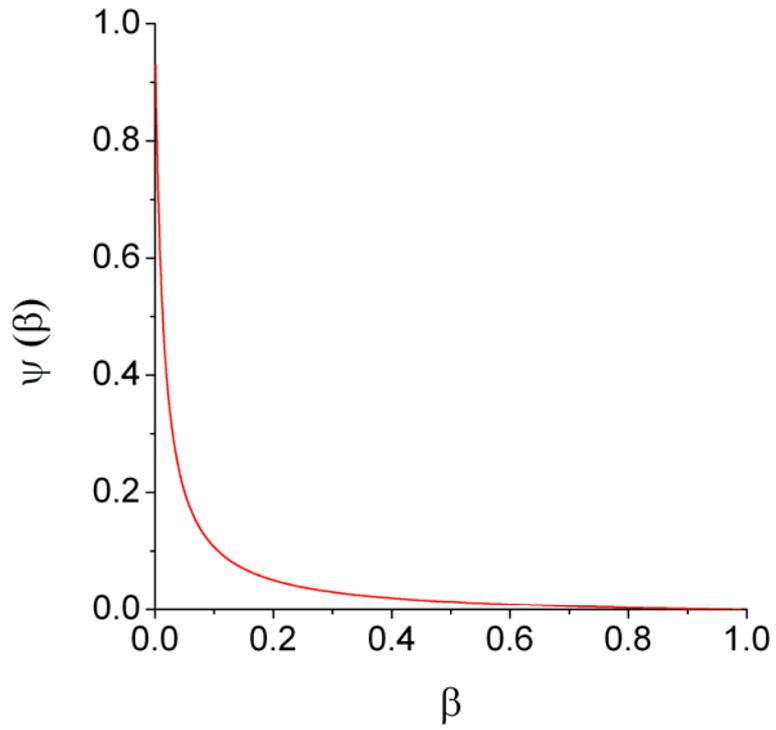

Fig. 1. The ratio $\psi = I/I_{iso}$ of the nucleation rate to the isothermal one as a function of the condensation coefficient $\beta$.

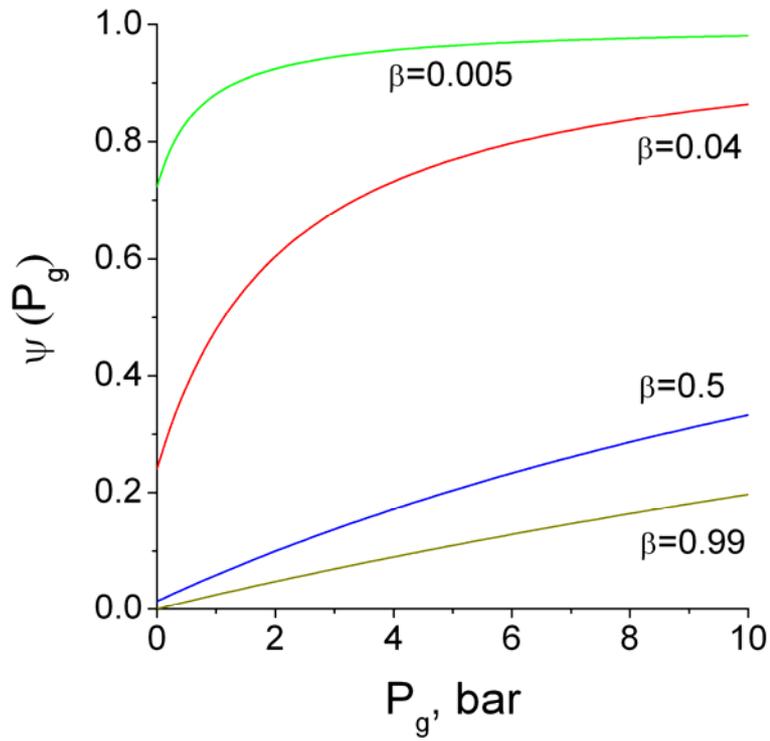

Fig. 2. The ratio $\psi = I/I_{iso}$ of the nucleation rate to the isothermal one as a function of the argon pressure $P_g$ for different values of the condensation coefficient $\beta$; the argon thermal accommodation coefficient $\beta_{\varepsilon,g}$ is put equal to $0.5$.

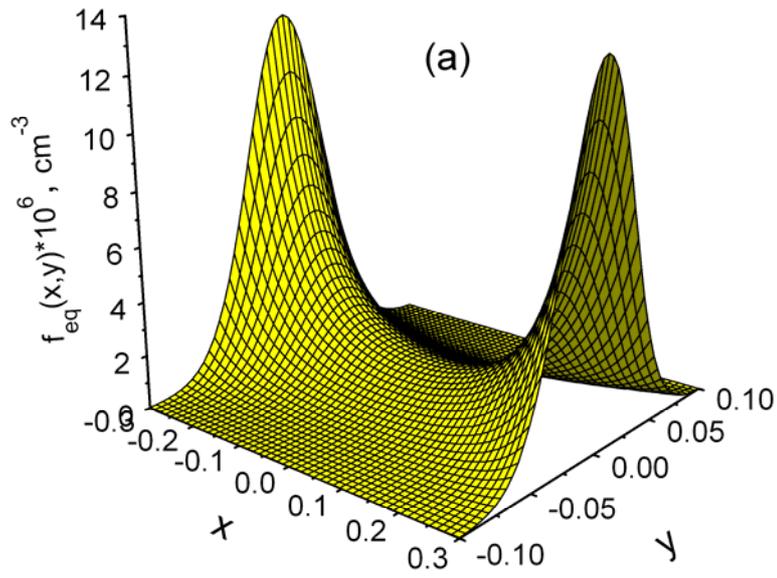

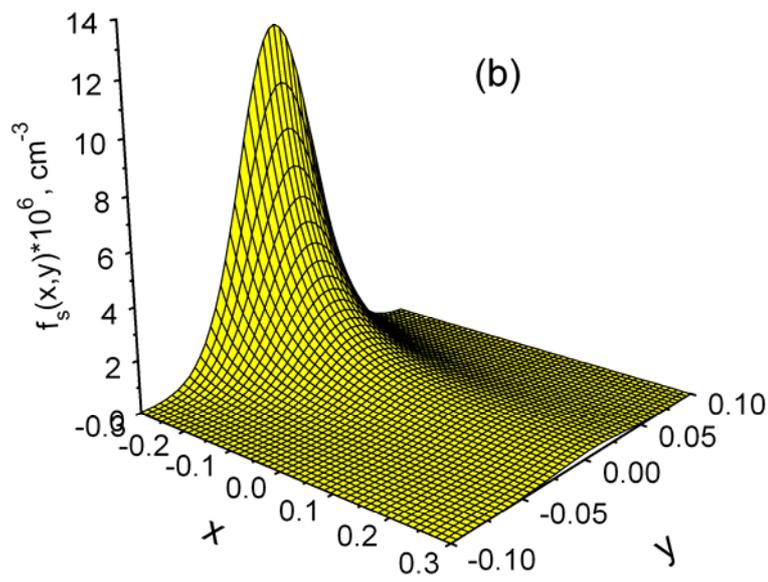

Fig.3. Equilibrium (a), eq. (101), and steady state (b), eq. (97), distribution functions of droplets in the $x$-region $\Omega = [-\Delta, \Delta]$, $\Delta = 0.3$, for the water vapor at the Table conditions.

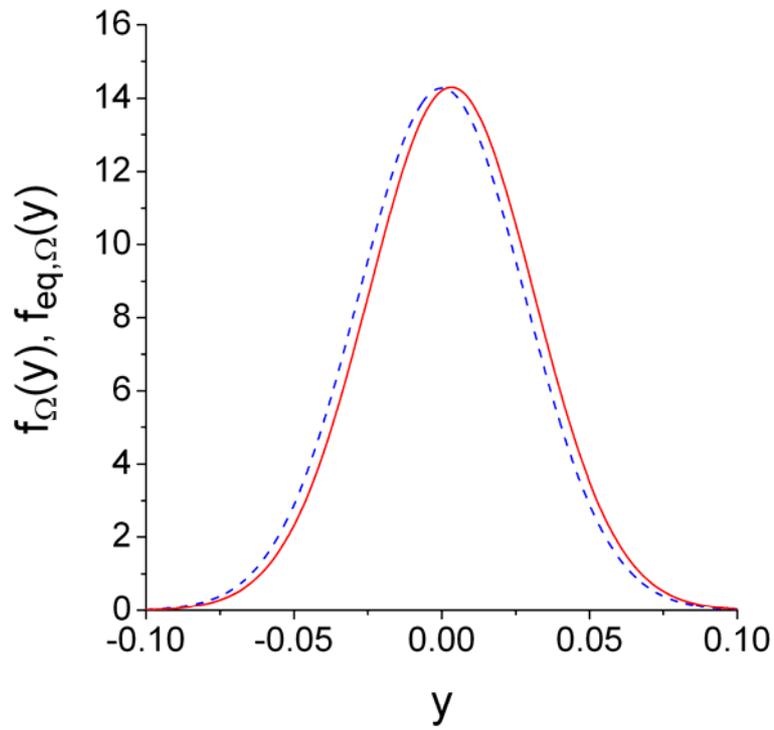

Fig. 4. The temperature distributions $f_\Omega(y)$ (steady state, solid) and $f_{eq,\Omega}(y)$ (equilibrium, dashed), eq. (102), of overcritical droplets in the region $\Omega = [0, \Delta]$ for the water vapor at the Table conditions.

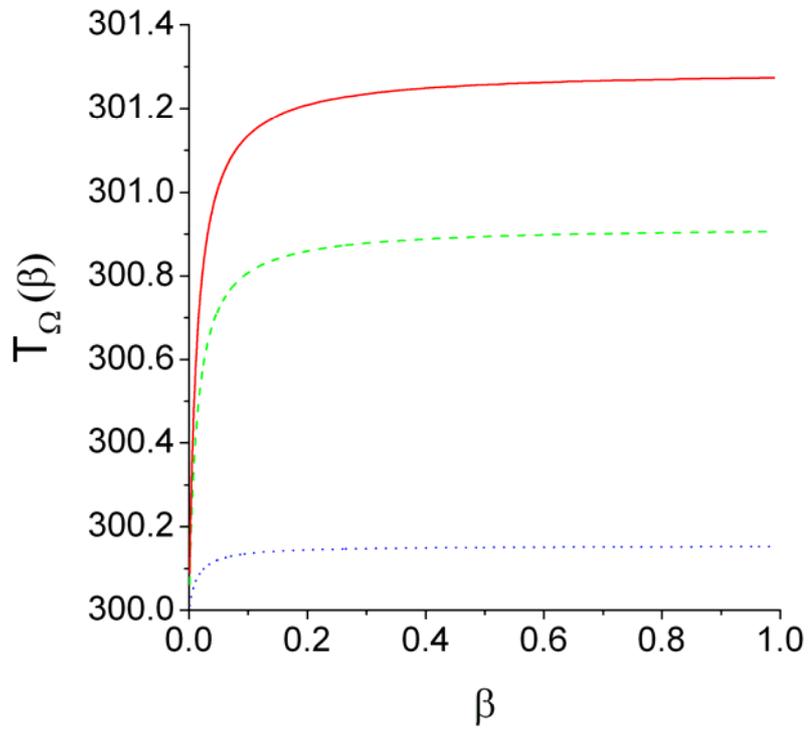

Fig. 5. The mean temperature of droplets, $\overline{T}_\Omega(\beta)$, as a function of the condensation coefficient for the regions $\Omega = [-\Delta, 0]$ (dotted), $\Omega = [0, \Delta/3]$ (dashed), and $\Omega = [0, \Delta]$ (solid).

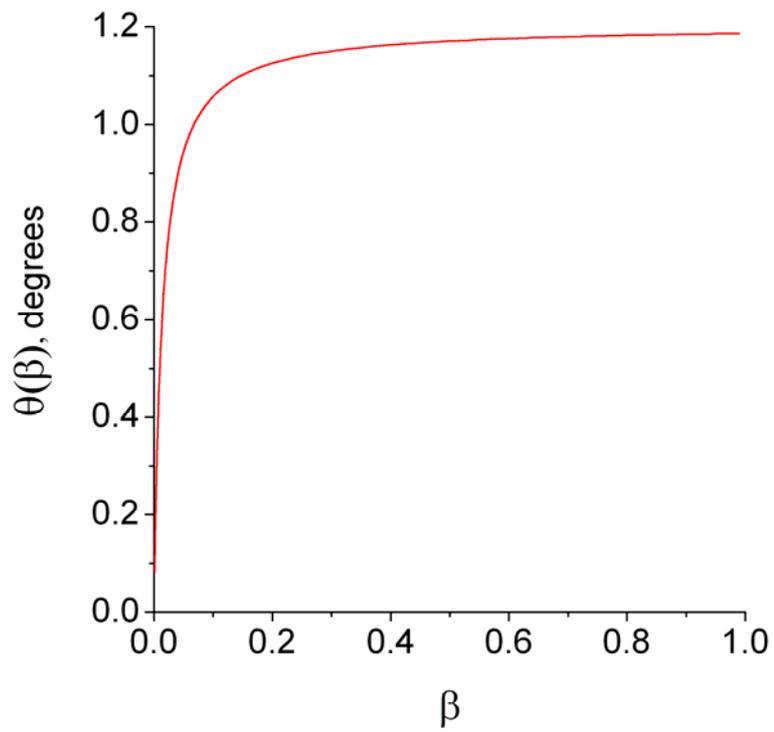

Fig. 6. Deflection of the flux of droplets from the steepest descent direction as a function of the condensation coefficient.

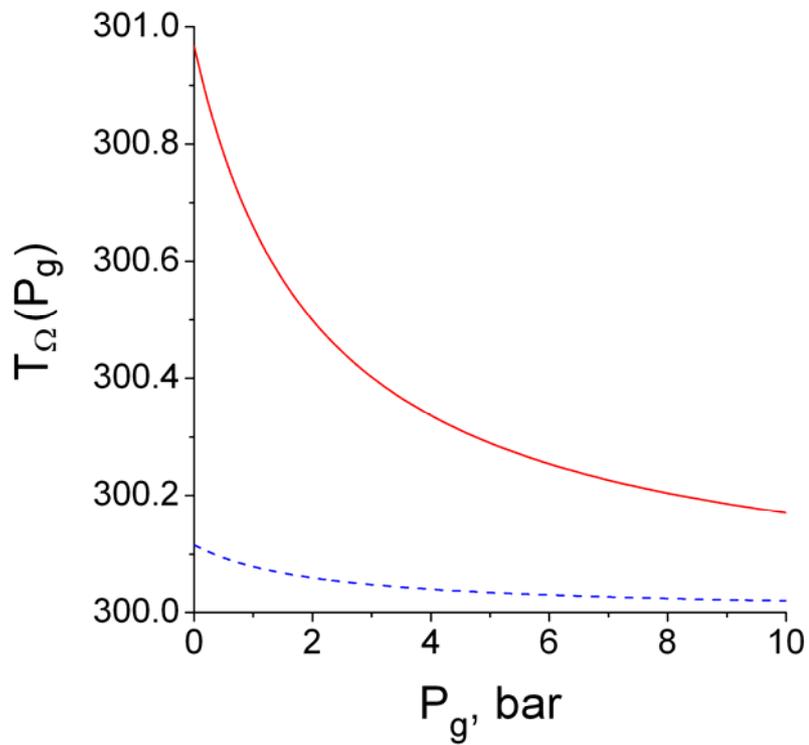

Fig. 7. Dependence of the mean temperature of the subcritical ($\Omega = [-\Delta, 0]$, dashed) and overcritical ($\Omega = [0, \Delta]$, solid) droplets in the water vapor at the Table conditions on the inert background gas (argon) pressure.